\documentclass[preprint2]{aastex}

\slugcomment{manuscript for AJ, vers.3, 100426}

\shorttitle{Brorfelde Schmidt CCD Catalog}
\shortauthors{Zacharias et al.}

\begin{document}

\title{Brorfelde Schmidt CCD Catalog (BSCC)}



\author{N. Zacharias$^1$,
        O. H. Einicke$^2$,
        K. Augustesen$^2$,
        J. V. Clausen$^2$,
        C. Finch$^1$,
        E. H{\o}g$^2$,
        G. L. Wycoff$^1$}

\email{nz@usno.navy.mil}

\affil{$^1$U.S.~Naval Observatory, 3450 Mass.Ave.~NW, Washington DC 20392;\\
       $^2$Niels Bohr Institute, C{\o}penhagen University,
           Juliane Maries Vej 30, DK-2100 C{\o}penhagen Oe, Denmark} 



\begin{abstract}
The Brorfelde Schmidt CCD Catalog (BSCC) contains about
13.7 million stars, north of $+49^\circ$ Declination
with precise positions and V, R photometry.
The catalog has been constructed from the reductions of
18,667 CCD frames observed with the Brorfelde Schmidt Telescope
between 2000 and 2007.  The Tycho-2 catalog was used for 
astrometric and photometric reference stars.
Errors of individual positions are about 20 to 200 mas
for stars in the R = 10 to 18 mag range.
External comparisons with 2MASS and SDSS reveal possible small systematic
errors in the BSCC of up to about 30 mas.
The catalog is supplemented with J, H, and K$_{s}$ magnitudes from the
2MASS catalog. 
\end{abstract}


\section{INTRODUCTION}

The Brorfelde Schmidt telescope \citep{bstel} was used with a 
2k by 2k charge-coupled device (CCD) camera \citep{bsccd}
to observe the sky from 
$+49.6^{\circ}$ to $+90^{\circ}$ declination, with additional
observations taken in selected fields north of $-10^{\circ}$.
The main idea of the project was to supplement the Carlsberg Meridian
Circle Catalog (CMC14) \citep{cmc14}, which covers the $-30^{\circ}$
to $+50^{\circ}$ declination area.
However, the resulting Brorfelde Schmidt CCD Catalog (BSCC) does 
not have a complete sky coverage north of $+49.6^{\circ}$,
with about 30\% of that area not observed.
The sky coverage of the BSCC is shown in Figure 1.
This catalog will be very valuable for determination of highly
accurate proper motions of faint stars when combining with future
data, as for example the upcoming USNO Robotic Telescope (URAT)
survey \citep{urat} and for assisting in systematic error analysis
of existing catalogs like the UCAC3 \citep{u3r} and PPMXL \cite{ppmxl}.

The CCD data were obtained during the
period of 2000 to 2007 and all applicable 18,667 object CCD frames
together with additional calibration frames were sent to the U.S.~Naval
Observatory (USNO) in 2009 for processing.
The USNO CCD Astrograph Catalog (UCAC) project \cite{u3r} reduction
pipeline \citep{u3x} was modified and used for this project.
A catalog of 13.7 million stars for the magnitude range
R = 10 to 18 has been constructed with precise positions on the 
International Celestial Reference System (ICRS) by use of the 
Tycho-2 \citep{tycho2} reference star catalog.

Photometry in V and R has been obtained for most of these stars
from the CCD data, with Two-micron All Sky Survey (2MASS)
near-infrared photometry added to the catalog \citep{2mass}.
The filters used at the Brorfelde Schmidt for this project are
approximating the standard Johnson V and Cousin R filters.

Details about the data acquisition and astrometric and photometric
reductions are presented here, followed by a description of the
resulting catalog.
External comparisons were performed with the Sloan Digital Sky
Survey (SDSS) release 7 data (www.sdss.org/DR7/) and the 2MASS 
point source catalog (www.ipac.caltech.edu/2mass).
The BSCC is available from the Strasbourg Data Center (CDS).

\section{OBSERVATIONS}

The observations were performed during 328 nights between
2000 February 13 and 2007 September 6 at the (now closed) 
50/77 cm Brorfelde Schmidt Telescope belonging to Copenhagen University.
It is located at 55$^{\circ}$ 37' 31" N, 11$^{\circ}$ 39' 59" E.
Its focal length, 150 cm, and thereby its focal plane scale of 137 arcsec/mm
was chosen to match the typical atmospheric seeing in Brorfelde of 2 arcsec
and match the resolution of available photographic emulsions.

Since 1999, the telescope has been equipped with a 2048x2048 thick,
front side illuminated Kodak KAF-4201 CCD, which is cooled by a
three-stage Peltier element, which again is water cooled \citep{bsccd}.
The pixel size of 9 microns yields a scale of 1.24 arcsec/pixel and
a field size of 42x42 arcmin.

For the present project, exposures were obtained through B, V, R, and I
filters, respectively, with transmission curves shown in Figure 2.
Exposure times were 60 to 600 sec (B), 10 to 600 sec (V,R), and 90 to 600 
sec (I), with the majority of data being 90 sec V and R images.
All exposures were taken close to the meridian with maximum deviations
of $-12^{\circ}$ and $+10^{\circ}$, respectively.
Bias and dark exposures were taken each night, and dome flat exposures
were secured at regular intervals during daytime.

\section{DATA REDUCTION}

All data reduction was performed in the Cataloging Division of the 
U.S.~Naval Observatory, using DS9 and IRAF for manual investigation
of FITS CCD frames and custom Fortran software for all pipeline processing,
based on the code used for the UCAC project.

\subsection{Pixel Data Processing}

Without assuming metadata to be correct, all available CCD data FITS files
were read with custom utility code to derive statistics about pixel value
histograms and to retrieve header information.
Visual inspection of the resulting output tables showed that in most cases
the first bias frame taken at the beginning of each night displays a large
number of bright pixels.  All first images of any given night were thus
excluded from the following processing.
Depending on the statistical properties of the CCD frames they were
identified and grouped into bias, darks, flat, and science object files.
A total of 25,190 CCD frames were provided by Copenhagen University,
of which 2,884 were identified as bias, 2,245 as dark, 271 as flat,
and 18,926 as science frames.  
A total of 60 frames failed the automated classification and manual
inspection revealed overexposed flats, which were disregarded.
The 804 binned data frames were also not processed.

Dark frames were utilized to identify ``bad" pixels which were flagged
and excluded from further processing.  
For each standard exposure time combined darks were derived.
For exposure times which were not used often a single master dark
was calculated from all available dark frames of that exposure time.
For the 20, 25, and 60 second exposures, master darks by observing
year were generated, while for the most often used 90 second exposures
a combined dark per month of observing was established.
A single bias-offset value per individual dark frame (added to all pixels
of a frame; a different value for different frames) was applied to bring 
all frames to the same bias level value before combining. 
However, individual frames
were not corrected for bias frame subtraction because the bias level
structure was found to be well corrected with applying combined darks.

A combined, master flat file was created for each filter and each
year of observing from the available dome flat data, excluding
overexposed frames.  The appropriate master dark frames were
subtracted from the flats before combining.  Individual flat frames
were normalized, low/high pixel values excluded,  and the mean taken
over the remaining values for each pixel to arrive at a combined flat.

All combined flats were analyzed to identify outlier pixel values
and largely variable pixels.  Those were added to the ``bad pixel" map
and not used in the astrometric and photometric reductions of stellar
images.

\subsection{Astrometric Reductions}

Science frames were grouped by filter and reduced separately.  
Tycho-2 reference stars were identified in the $x,y$ data and
weighted, least-squares, conventional ``plate" adjustments (CPA) 
run on all applicable frames.  A total of 127 I, 9350 R, 9115 V, 
and 75 B filter frames provided successful solutions (18,667 CCD
frames all together).
Of these, a total of 1,888 CCD frames are outside the $+49.6^{\circ}$
to $+90^{\circ}$ area which constitutes the published catalog data,
thus 16,779 CCD frames were used for the BSCC.
The number of reference stars used per CCD frame is shown in the
histogram of Fig.~3.  There are 427 frames with 9 or less reference
stars in the area north of $+49.5^{\circ}$, while the typical number
of Tycho-2 reference stars used per frame is about 12 to 35,
with a few frames up to about 100.

A linear plate model (6 parameters) was adopted without correcting
for differential refraction or aberration.  Also no third order
geometric distortion was pre-applied, which is inherent in Schmidt 
telescope imaging (curved focal plane and tangential projected 
$\xi, \eta$).  These procedures are adequate due to the small
field of view and the presence of other distortions for example
from the filters.  However, the combined effect of all geometric 
distortions is determined empirically as follows, and has been
corrected.

A total of 312,522 and 292,957 residuals (with 2 coordinates each)
were available from the astrometric reductions of all R and V filter
observations, respectively.
The residuals were stacked up in bins as function of focal plane
$x,y$ coordinates and slightly smoothed by weighted average with
neighboring bins.  The resulting field distortion patterns (FDP)
are shown in Fig.~4.  Linear interpolation between the bins were
performed to arrive at the FDP correction values which were
applied to the $x,y$ data prior to the following iteration of the
CPA reduction. 

Residuals of the final CPA reductions were plotted as a function
of $x,y$ coordinate, radial distance from the frame center,
magnitude, color, and coma term (product of magnitude and coordinate).
The largest systematic errors, up to about 20 mas, were found as a 
function of magnitude (Figs.~5, 6), with root-mean-square (RMS) scatter 
shown in Fig.~7.  For most of these 90 sec exposures in the survey
saturation is around R=10 preventing from reaching even smaller
positional errors at the bright end.  Errors for stars fainter
than 10th magnitude are dominated by the reference star contribution.
A complex, non-linear dependence for the $y$ coordinate 
(declination) as a function of coma-x and coma-y is found (not shown)
with amplitudes up to 20 mas.
These systematic errors are at the limit of the reference star
catalog at the epoch of the CCD observations.  No further
investigation was performed and no attempt to correct the derived
star positions for any such magnitude dependent systematic errors
was made.

The final catalog positions are obtained from a weighted mean of
all individual observations for each star.  Objects matching within
2.0 arcsec are assumed to be the same star.  
Thus the positions provided in this observational catalog are based 
on the mean of the V and R band observations, given at mean epoch of
observation, on the ICRS, as represented by the Tycho-2 catalog.

\subsection{Photometric Reductions}

Differential photometric reductions were performed with respect to
Tycho-2 standards on individual CCD frames. 
Instrumental magnitudes were derived from the integral (volume) of
the fitted stellar image profiles above the local background using
the modified UCAC reduction pipeline while performing the 
astrometric pixel reduction step.
In the following, only Tycho-2 stars with ``good" photometry flag,
brighter than V = 12.5, and not saturated on CCD frames were used.
Furthermore, stars with (B$-$V) $\le$ $-$0.2 and (B$-$V) $\ge$ 1.5
were excluded and stars in the (B$-$V) range of 1.1 to 1.5 were used
with reduced weight.

A magnitude zero-point ($V_{0}$) per V-band CCD frame was fitted to 
transform the instrumental magnitudes ($V_{i}$) into standard V 
magnitudes ($V_{s}$), using Tycho-2 references for $V_{s}$ in
a weighted least-squares adjustment according to

\[  V_{0} \ = \ V_{s} - \ V_{i} \].

Similarly,

\[  R_{0} \ = \ R_{s} - \ R_{i} \]

was determined using B and V magnitudes from Tycho-2 as standards
with the approximation

\[  R_{s} \ \approx \ V_{s} \  - \ s \ (B - V)  \]

The color slope term, $s = 0.45$, was derived from linear fits to 
sample CCD data with many reference stars and solving for both the 
slope and constant term.  The average resulting slope term was adopted
and used to solve only for the photometric zero-point constant
in all R-band CCD frames. 

Fig.~8 shows the distribution of the standard error of this
photometric constant.  For most frames this error is 20 to 40 millimag.
The absolute photometric error of stars in the BSCC has to be larger
than this photometric constant error, while the internal, photometric
precision is better for well exposed stars.

\section{RESULTS AND EXTERNAL COMPARISONS}

\subsection{The Catalog}

The BSCC contains 13,771,775 stars north of $+49.6^{\circ}$ declination
and is sorted by declination.  Of these, 13,157,292 do have a match
with a 2MASS star within 2 arcsec, and J, H, K$_{s}$ photometry with errors
were copied from the 2MASS into BSCC.  Stars based on a single CCD
observation and not matched with 2MASS did not enter the released catalog.
Fig.~9 shows the distribution of stars in the BSCC by R and V magnitude.
Completeness is expected up to about R = 17.5 with a limiting magnitude
of about R = 19.
There are 583,043 stars in the catalog without R magnitude, and
4,311,081 do not have a V magnitude. 

The BSCC data file is 1.7 GB, formatted, ASCII.  Some sample
lines are listed in Table 1, with the data format explained in
Table 2.  This is an observational catalog of mean positions at
a mean epoch, which is slightly different for each star.
The positions are on the International Celestial Reference System
(ICRS) by means of the Tycho-2 reference star catalog.
There are no proper motions provided.
Stars are identified by the IAU registered acronym "BSCC", followed
by a fixed-length, 8-digit running record number without a space. 

Fig.~10 shows the mean astrometric errors (per coordinate) of the 
BSCC as a function of R magnitude.  The filled circles show the
errors as derived from the model (including $x,y$ fit precision
and expected contribution from the ``plate" adjustment solution),
while the open squares show the error from the observed scatter of 
individual positions.  These are small number statistics for
individual stars (with typically 2 to 4 images) but become
meaningful when averaged over many stars (for the R $\ge$ 9 range).
Many stars in the 10 to 15 mag range have internal errors of
about 20 mas in Dec and about 25 mas in RA.
A near saturation effect seems to be present for stars brighter
than R = 11 mag showing a slightly larger, observed, scatter
error than at R = 11 mag.

Fig.~11 shows the formal, photometric errors in the BSCC R and V
magnitudes derived from the scatter of individual magnitudes.
Data shown for stars brighter than about 9th mag are affected by
small number statistics.  Systematic errors are not included here.
The precision of the photometry is on the 3 to 5 \% level for the
10 to 15 mag range, then increasing according to the lower 
signal-to-noise ratio to about 0.15 mag at R = 18.

\subsection{Astrometric Comparisons}

Unweighted position differences of BSCC minus Tycho-2, averaged
over 100 stars per dot are shown in Fig.~12 as a function of magnitude.
Tycho-2 proper motions have been applied to bring the positions to
the epoch of individual BSCC stars.
The small overall offset in declination is caused by the weighted
adjustment in the CCD reductions of individual frames combined 
with the magnitude dependent pattern shown in Fig.~6.
The differences in RA average to about zero, as expected.

Position differences of BSCC with respect to 2MASS are shown in Fig.~13.
No proper motions have been applied because neither the 2MASS nor the
BSCC do have proper motions.  
The epoch difference between individual BSCC and 
2MASS observations of a star is within $\pm$ 3 years.
Over 11 million stars were matched.
There is a small systematic difference along RA increasing toward
faint stars reaching about 30 mas at R=18, while for Dec there is
a remarkable consistency between the BSCC and 2MASS data with
only about 10 mas differences at the very faint end at R=18
and systematic differences less than about 5 mas for the entire 
magnitude range of R = 10 to 17.

Fig.~14 shows the position differences BSCC$-$2MASS as a function
of declination.  CCD frame size patterns are clearly seen with
systematic position differences of up to about 50 mas. 
Some of this is caused by not having proper motions applied.
The break at $\delta = 60^{\circ}$ is explained by the mean
epoch of BSCC observations (Fig.~15).  Data for $\delta < 60^{\circ}$
were observed, on average, at a much later epoch than the other data,
resulting in a larger epoch difference to 2MASS data.
This in turn gives larger position differences because no
proper motions are applied.
The data for $\delta \ge 78^{\circ}$ were observed at a mid-range
epoch, leading also to an increased scatter in the position
differences as compared to data between $60^{\circ}$ and $78^{\circ}$.
Fig.~16 shows the BSCC$-$2MASS position differences as a function
of right ascension.  The average offset for the RA component is
a result from the magnitude equation seen earlier (Fig.~13).
In addition, a sine wave pattern of about 20 mas amplitude is 
present in either coordinate.  This could be caused by galactic 
dynamics due to the lack of proper motions in this comparison.

A representative subset of the Sloan Digital Sky Survey (SDSS)
release 7 data \citep{sdss7}, \citep{sdssPM}
was selected ($200^{\circ} \le \alpha \le 270^{\circ}$,
$+58^{\circ} \le \delta \le +90^{\circ}$).
A match with BSCC resulted in 49,692 common stars and SDSS proper
motions were applied.
The position differences BSCC$-$SDSS are shown in Figs.~17 and 18 as 
functions of magnitude and color, respectively.
The RA coordinate shows a small magnitude equation, about 5 mas/mag,
similar to the BSCC$-$2MASS differences (Fig.~13), for the 14 to 18 mag
range.  The Dec coordinate displays a constant offset of about 10 mas
over that magnitude range.
There is no dependence of the BSCC$-$SDSS position differences as
a function of color, except for the average offset mentioned before.

Fig.~19 shows the position differences BSCC$-$CMC14 as a function
of R magnitude.  No proper motions were applied and the epoch 
difference is only a few years.
Note that the overlap between these 2 catalogs is
very small, involving only 194 CCD frames taken at $+50^{\circ}$ 
declination in the RA = 8 to 16 hour range with 20,596 stars in common.
Fig.~20 shows the BSCC$-$2MASS position differences for the same
area in the sky as used for the CMC14 comparison, which is significantly
different than the average BSCC$-$2MASS differences of Fig.~12.
Stars with position differences over 300 mas in either coordinate
were excluded for this comparison.
For the RA component, CMC14 agrees with 2MASS but both are offset
with respect to BSCC by about $-30$ mas.
For the Dec component, BSCC agrees with 2MASS but CMC14 is offset
by about +30 mas with respect to the other 2 catalogs.

\subsection{Photometric Comparisons}

Fig.~21 shows the difference in BSCC R minus CMC14 r magnitude
as a function of CMC14 r magnitude.  A narrow distribution is
seen with systematic differences in the range of about $\pm$0.1 mag.
Fig.~22 shows the color-color diagram between the BSCC V, R and
SDSS g and r bandpasses. 
The BSCC R band magnitudes match the SDSS r magnitudes very
well for (g$-$r) colors between 0.2 and 1.0, with significant
systematic differences for redder stars.
The BSCC V and SDSS g magnitudes are significantly different
(up to 0.4 mag) for all colors, as expected.

\section{DISCUSSION}

Considering the low resolution (1.24 arcsec/pixel) of the Brorfelde
Schmidt survey data the catalog is amazingly accurate, with repeatability
of observations (from overlapping fields) of about 20 mas per coordinate
for well exposed stars.
Systematic errors in observed star positions are also small, on the 30 mas
level or below, with particularly well controlled systematic errors as
a function of magnitude (see for example the comparison with 2MASS data).

However, these survey data could benefit from a denser reference star
catalog than the Tycho-2.  Most of the highly accurate Tycho stars are
just overexposed in the Schmidt CCD data, while the accessible Tycho
stars around magnitude 11 to 12 are affected by relatively large errors.
Typical CPA solutions have a standard error of 60 to 80 mas (combined
error of the $x,y$ data and the reference star positional errors,
per star and coordinate).
Thus the ``plate" parameters are not very well defined, which leads
to significant scatter in samples of a few CCD frames, as seen for
example in the comparison of BSCC data with the CMC14 catalog which
barely overlap at declination $+50^{\circ}$.
A block-adjustment type solution for overlapping frames is difficult
with the BSCC data due to the large gaps in the survey area coverage
and not enough overlap between adjacent CCD images (no 2 or 4-fold
overlap pattern).

Similarly the photometric results suffer from a weak determination
of the zero-point per CCD frame due to the small number of available,
highly accurate standard stars.  Once a denser photometric catalog
than the Tycho-2 becomes available, the BSCC data would benefit from
a re-reduction.

The most important aspect of the BSCC astrometric data is the high
signal-to-noise data for stars at the faint UCAC end,
i.e.~stars around R = 15 to 16.  The precision of the BSCC positions
is about 40 mas per coordinate, while the UCAC data are on
the 70 mas level at those magnitudes.
However, only a small area of the sky could be covered with this
7-year effort. Full-sky coverage with high accuracy down to about 
R = 17.5 will hopefully be achieved soon with the URAT project
\citep{urat}.  Combining URAT data with the BSCC will provide
excellent proper motions for those faint stars on the 3 mas/yr
level, completely based on CCD observations for early and current
epoch data.

\acknowledgments

We wish to thank the former director of the Copenhagen University
Observatory, Henning J{\o}rgensen, who played a key role in the
definition and initialization of the project.

This publication makes use of data products from the Two Micron
All Sky Survey, which is a joint project of the University of
Massachusetts and the Infrared Processing and Analysis Center/California
Institute of Technology, funded by the National
Aeronautics and Space Administration and the National Science
Foundation.

Funding for the SDSS and SDSS-II has been provided by the Alfred P. Sloan 
Foundation, the Participating Institutions, the National Science Foundation,
the U.S. Department of Energy, the National Aeronautics and Space 
Administration, the Japanese Monbukagakusho, the Max Planck Society,
and the Higher Education Funding Council for England. The SDSS Web Site 
is http://www.sdss.org/.

National Optical Astronomy Observatories (NOAO) is acknowledged for
IRAF, Smithonian Astrophysical Observatory for DS9 image display software,
and the California Institute of Technology for the {\em pgplot} software.



\clearpage



\begin{figure}
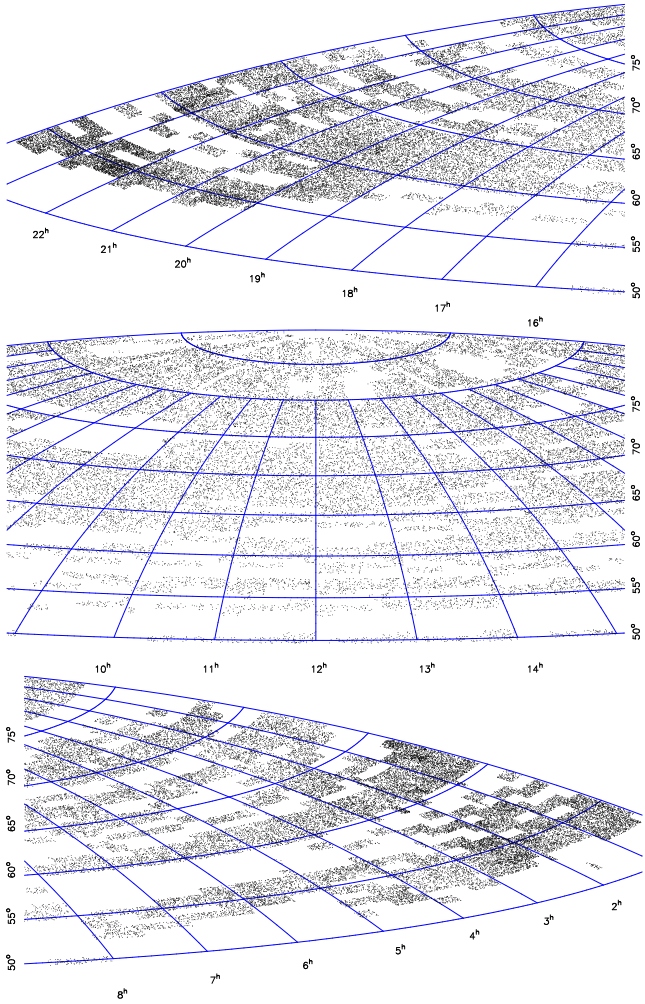
    
\epsscale{1.00}
\includegraphics[angle=-90,scale=.33]{fig01a.ps}
\includegraphics[angle=-90,scale=.33]{fig01b.ps}
\includegraphics[angle=-90,scale=.33]{fig01c.ps}
\caption{Distribution of BSCC stars on the sky.  
  Shown are stars in the R-mag range of 10 to 12 between 
  +49 and +90 declination in 3 sections split by RA on
  equal-area projections.}
\end{figure}

\begin{figure}   
\includegraphics[angle=0,scale=0.28]{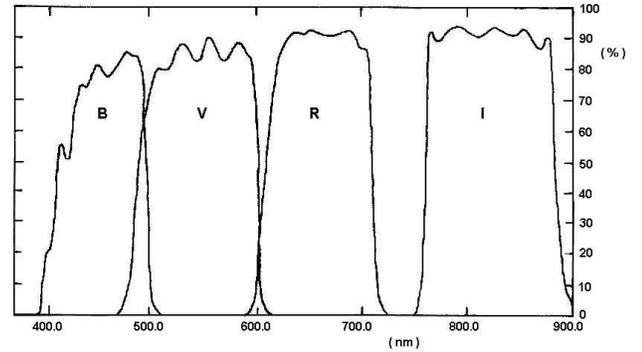}
\caption{Bandpass data for the B, V, R, and I filters
  used at the Brorfelde Schmidt telescope for this project.}
\end{figure}

\begin{figure}   
\includegraphics[angle=-90,scale=.33]{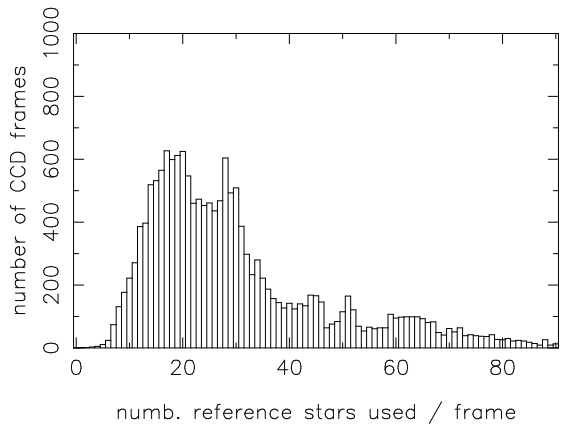}
\caption{Histogram of the number of reference stars
  (from Tycho-2) actually used per CCD frame in the
  astrometric reductions of the BSCC.}
\end{figure}

\begin{figure}   
\epsscale{1.00}
\includegraphics[angle=0,scale=.45]{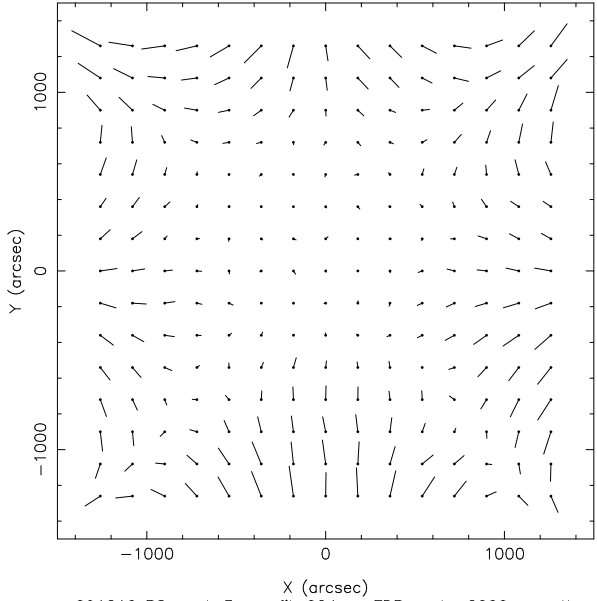}
\includegraphics[angle=0,scale=.45]{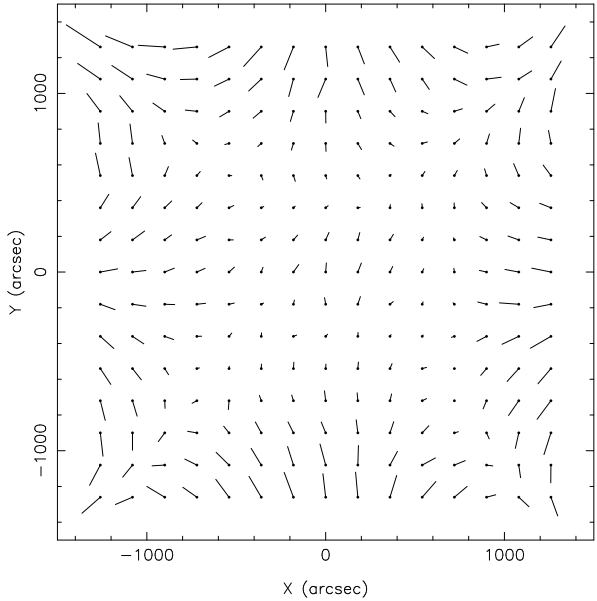}
\caption{Field distortion pattern derived from the
  residuals of Tycho-2 reference stars for all applicable
  R-band (top) and V-band (bottom) Brorfelde Schmidt CCD
  frame observations.  The scale of the residual vectors
  is 2000, thus the largest vectors are about 100 mas.}
\end{figure}

\begin{figure}  
\epsscale{1.00}
\plotone{fig05.ps}
\caption{Residuals from final reductions of R-band  
  observations with respect to Tycho-2 reference
  stars as a function of magnitude.  
  One dot represents the mean over 400 stars.}
\end{figure}

\begin{figure}  
\epsscale{1.00}
\plotone{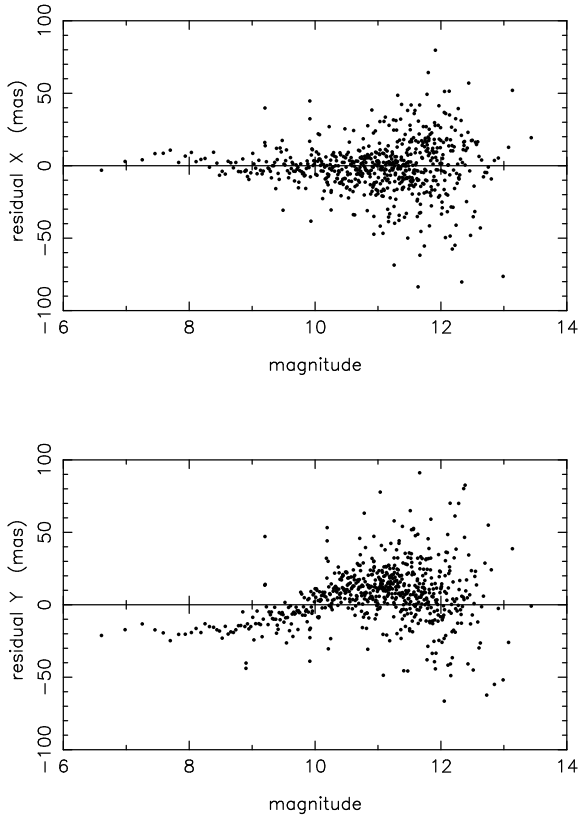} 
\caption{Same as the previous figure but for the V-band 
  observations.}
\end{figure}

\begin{figure}  
\epsscale{1.00}
\plottwo{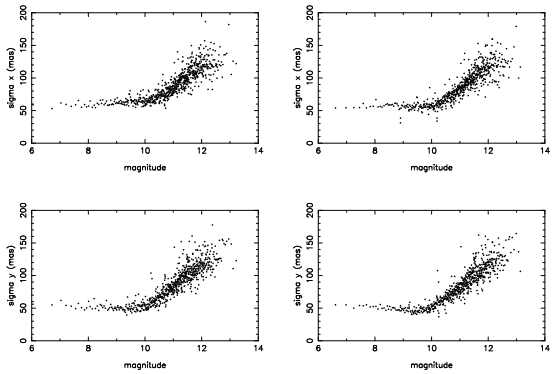}{fig07b.ps}
\caption{Scatter (standard deviation) of the residuals for
  R-band (left) and V-band (right) observations as a function 
  of magnitude.  One dot represents the mean over 400 stars.}
\end{figure}

\begin{figure}  
\includegraphics[angle=-90,scale=.35]{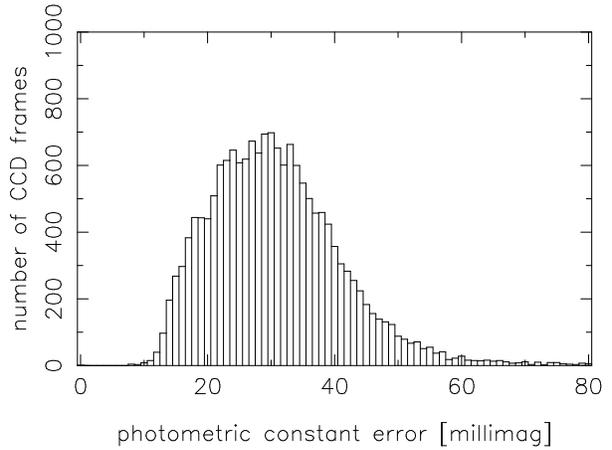}
\caption{Distribution of the standard error in determining
  the photometric constant between instrumental and absolute
  magnitudes.  This error varies significantly depending on
  the number and quality of Tycho-2 photometric standard
  stars available in any given CCD frame.}
\end{figure}

\begin{figure}  
\epsscale{1.00}
\plotone{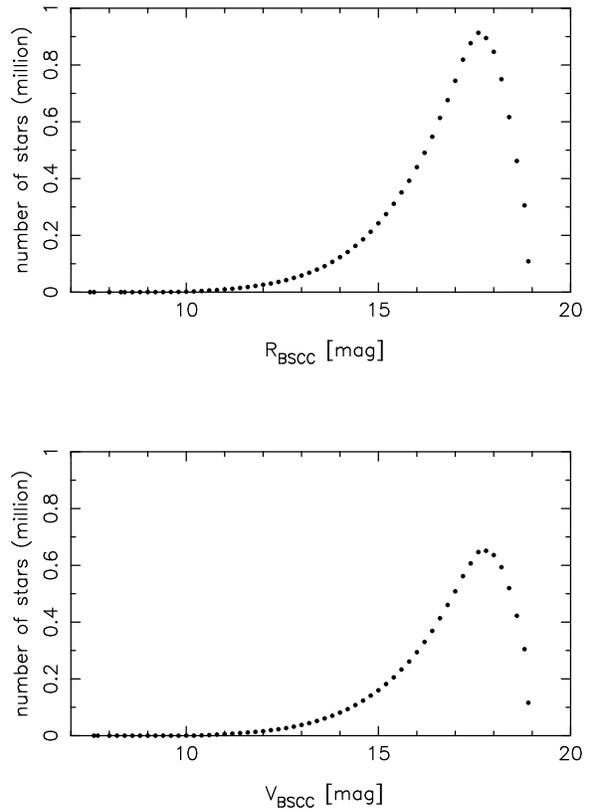}
\caption{Distribution of BSCC stars by R and V magnitude.} 
\end{figure}

\begin{figure}  
\epsscale{1.00}
\plotone{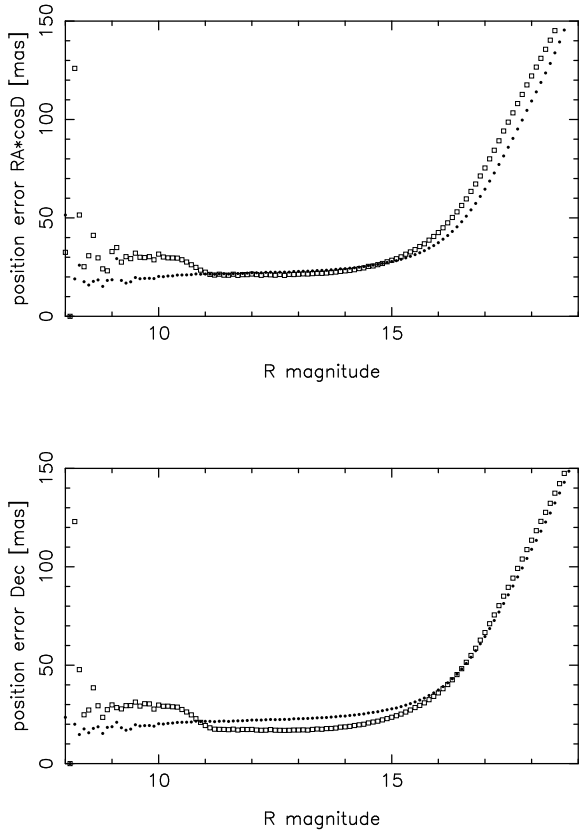} 
\caption{Mean positional errors as a function of R magnitude
  for the RA (top) and Dec (bottom) components.  The filled
  circles show the model standard error, while the open squares
  show the mean standard error from the scatter of individual 
  positions which contribute to a mean catalog position.}
\end{figure}

\begin{figure}  
\includegraphics[angle=-90,scale=.33]{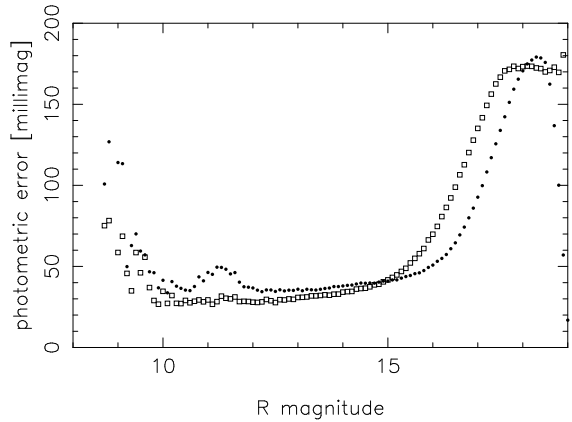}
\caption{Formal error on photometry of the BSCC as a function
  of R magnitude.  The filled circles are for the R mag error,
  while the open squares are for the V mag error.}
\end{figure}



\begin{figure}   
\epsscale{1.00}
\plotone{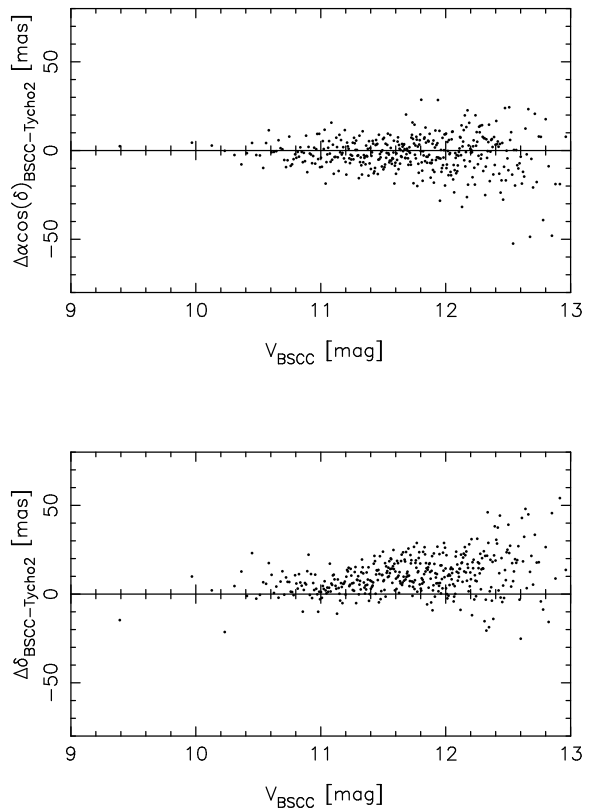}
\caption{Unweighted position differences BSCC$-$Tycho2 in 
  right ascension (top) and declination (bottom) as a function 
  of V magnitude. 
  One dot represents the mean over 200 stars.}
\end{figure}

\begin{figure}   
\epsscale{1.00}
\plotone{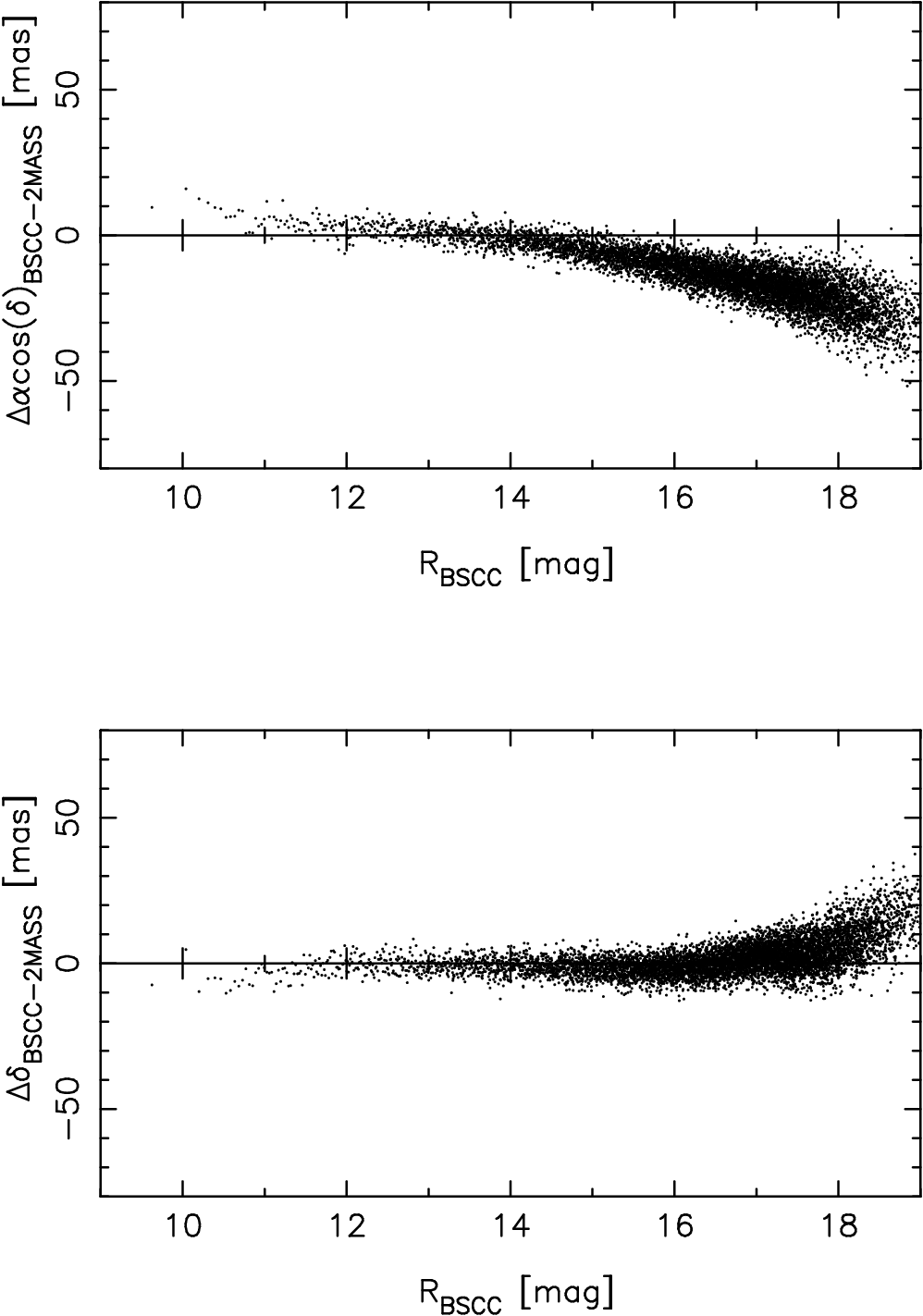} 
\caption{Position differences BSCC$-$2MASS in right ascension (top)
  and declination (bottom) as a function of R magnitude. 
  One dot represents the mean over 1500 stars.} 
\end{figure}


\begin{figure}   
\epsscale{1.00}
\plotone{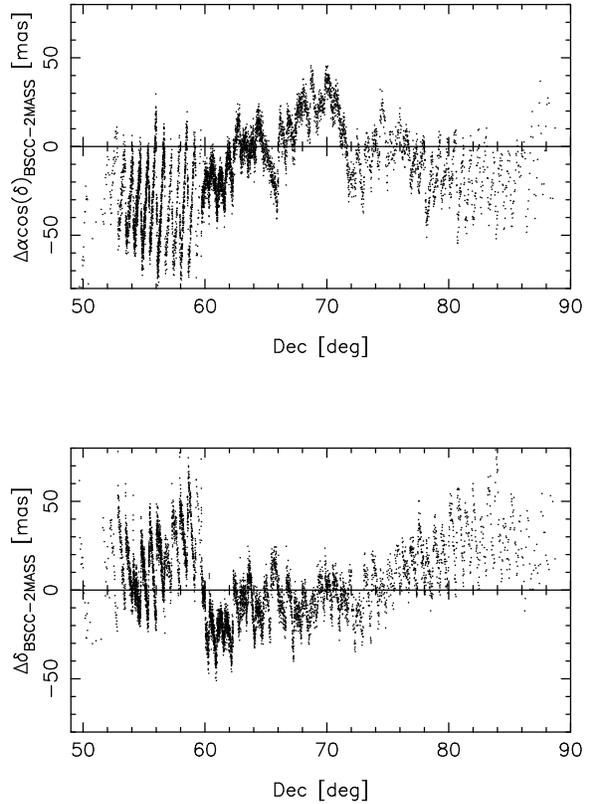} 
\caption{Position differences BSCC$-$2MASS in right ascension (top)
  and declination (bottom) as a function of declination. 
  One dot represents the mean over 1500 stars.} 
\end{figure}

\begin{figure}   
\includegraphics[angle=-90,scale=.33]{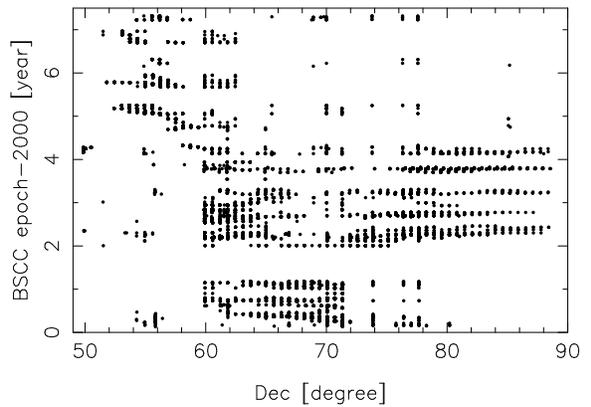}
\caption{Distribution of BSCC observation epochs as a
  function of declination.}
\end{figure}

\begin{figure}   
\epsscale{1.00}
\plotone{fig16.ps} 
\caption{Position differences BSCC$-$2MASS in right ascension (top)
  and declination (bottom) as a function of right ascension. 
  One dot represents the mean over 1500 stars.} 
\end{figure}

\begin{figure}   
\epsscale{1.00}
\plotone{fig17.ps}
\caption{Position differences BSCC$-$SDSS in right ascension (top)
  and declination (bottom) as a function of BSCC R magnitude. 
  One dot represents the mean over 200 stars.} 
\end{figure}

\clearpage

\begin{figure}   
\epsscale{1.00}
\plotone{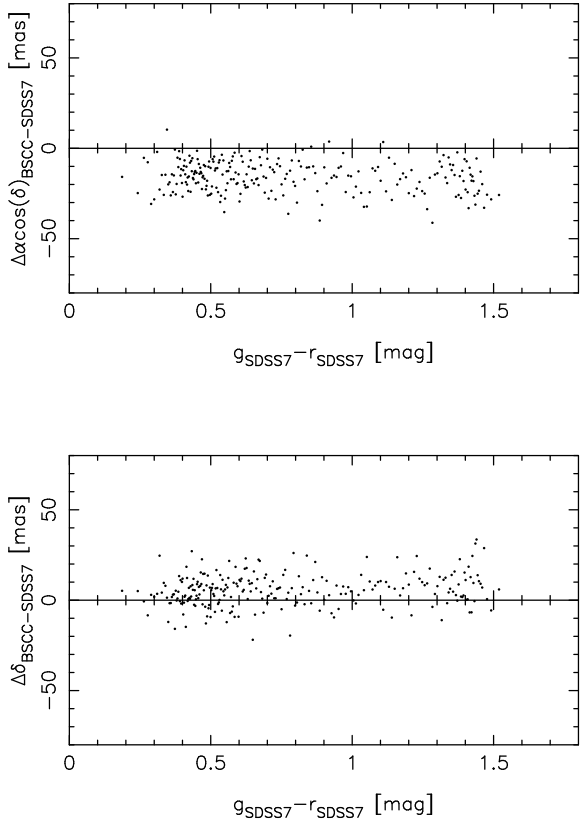}
\caption{Position differences BSCC$-$SDSS in right ascension (top)
  and declination (bottom) as a function of color. 
  One dot represents the mean over 200 stars.} 
\end{figure}



\begin{figure}    
\epsscale{1.0}
\plotone{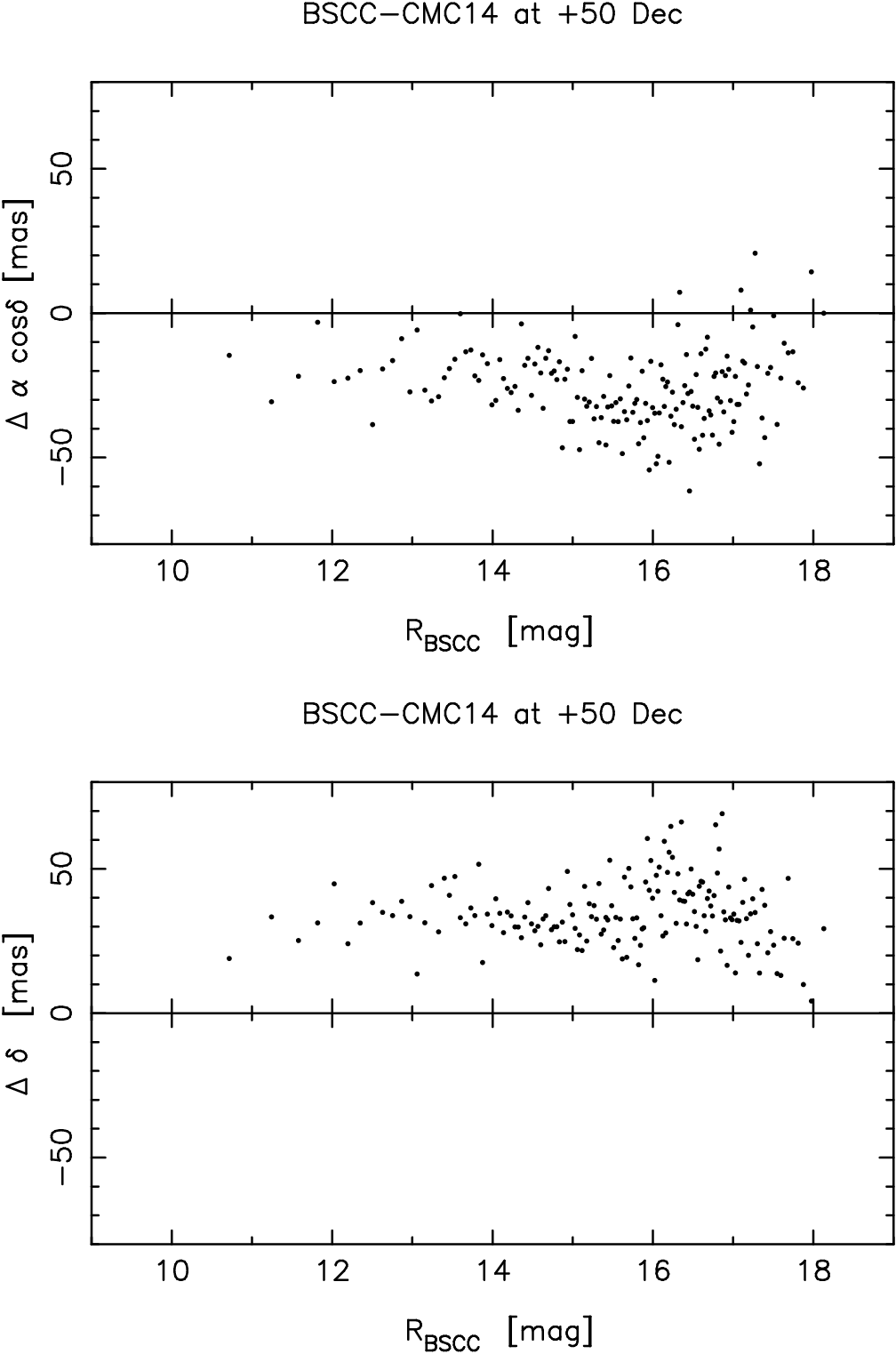}
\caption{Position differences BSCC$-$CMC14 for RA (top)
  and Dec (bottom) coordinate as a function of BSCC R-magnitude.
  One dot represents the mean over 100 stars.}
\end{figure}

\begin{figure}    
\epsscale{1.0}
\plotone{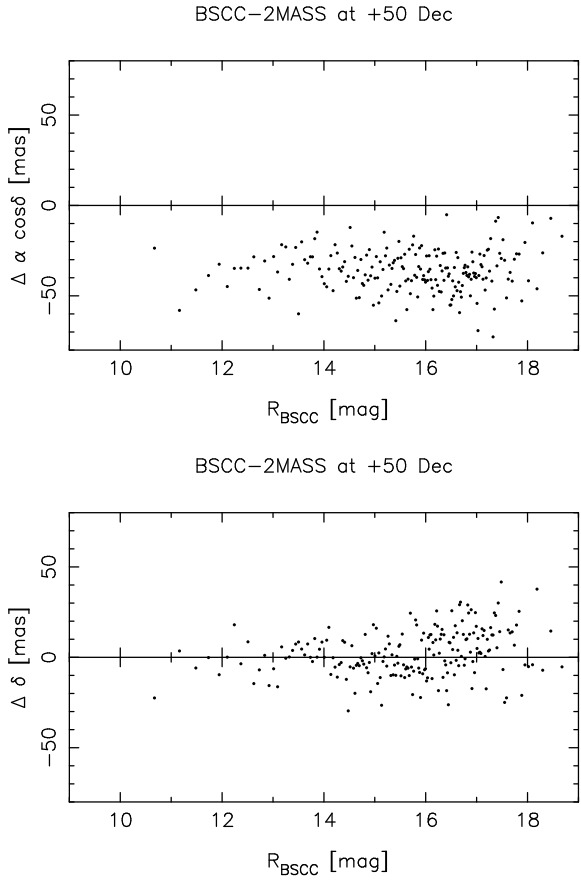}
\caption{Position differences BSCC$-$2MASS for RA (top)
  and Dec (bottom) coordinate as a function of BSCC R-magnitude,
  for the BSCC vs.~CMC14 overlap area only.
  One dot represents the mean over 100 stars.}
\end{figure}


\begin{figure}    
\epsscale{0.80}
\includegraphics[angle=-90,scale=.33]{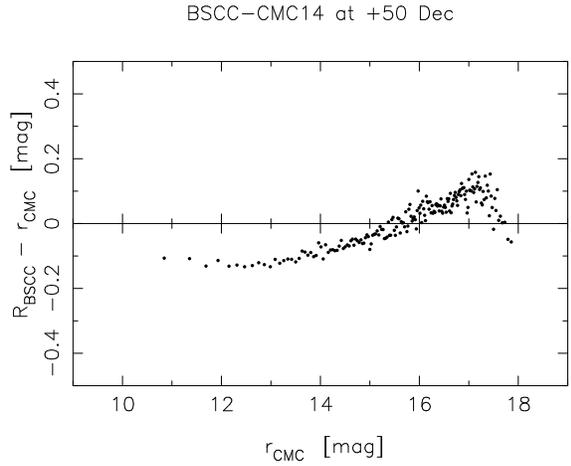}
\caption{Magnitude differences BSCC R $-$ CMC14 r 
  as a function of CMC14 r magnitude. 
  One dot represents the mean over 100 stars after
  excluding outliers.} 
\end{figure}

\begin{figure}
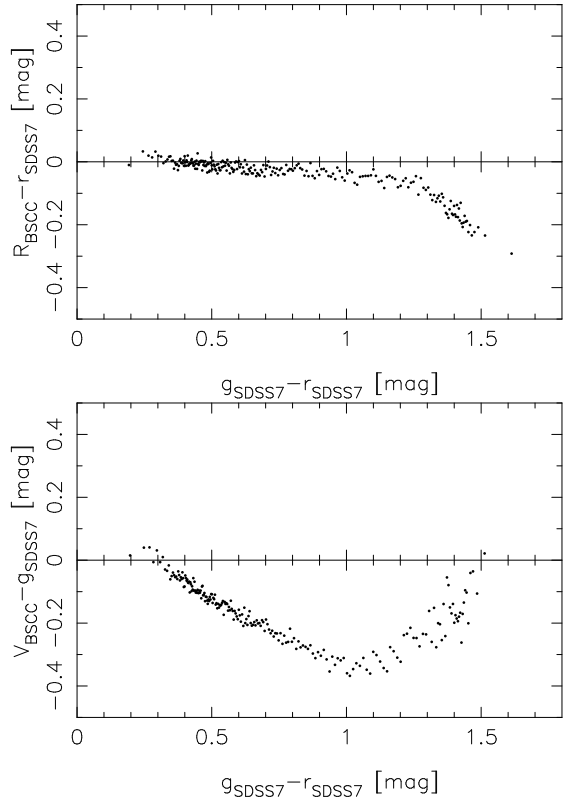
    
\epsscale{1.00}
\includegraphics[angle=-90,scale=.33]{fig22a.ps}
\includegraphics[angle=-90,scale=.33]{fig22b.ps}
\caption{Color-Color diagrams for BSCC R magnitude (top) and V magnitude
  (bottom) as compared to SDSS r and g magnitudes.
  One dot represents the mean over 200 stars after excluding outliers.}
\end{figure}





\clearpage

\begin{table}
\caption{The first 10 lines of the BSCC data file.}
\footnotesize
\begin{verbatim}
column    1         2   3   4   5   6 7 8 9   10 11 12    13 14 15    16  17 18         19    20  21    22  23    24 25
-----------------------------------------------------------------------------------------------------------------------
  688580578 178689890 205  81 205  81 1 1 0 4276  0  0 15424 -1  1 20000  -1  0  586589348 13675  25 13319  23 13305 34
  447877328 178690199 215 235  79  81 2 2 0 4251  0  0 16728 -1  1 17872  -1  1  789094466 15231  49 15146  82 14945 11
  455055784 178691095  80 280  63  62 2 2 0 4251  0  0 15082 -1  1 16426  -1  1  795487085 13414  26 13029  32 12972  3
  734522546 178691442 313  59 122 121 2 2 0 4172  0  0 14531 -1  1 15560  -1  1  663675738 13694  27 13343  32 13357  4
  455012579 178693216  12 107  56  56 2 2 0 4251  0  0 16377 -1  1 16782  -1  1  795487088 15153  43 14744  59 14719 10
  561019952 178694873  35  31  79  81 2 2 0 4172  0  0 15981 -1  1 16741  -1  1  479162850 15029  34 14515  54 14470  8
  727523129 178695017 229 229 229 229 1 1 0 4172  0  0 18172 -1  1 20000  -1  0 1028367395 15287  34 14778  52 14460  6
  433962337 178695446  30  88  61  61 2 2 0 4251  0  0 16496 -1  1 16866  -1  1  480839919 15330  55 14788  72 14807 10
  455332559 178695609  61  66  51  52 2 2 0 4251  0  0 15335 -1  1 16029  -1  1  795487090 14271  30 13871  46 13938  5
  440856487 178696770  14  24  53  52 2 2 0 4251  0  0 14425 -1  1 14782  -1  1  480944299 13567  26 13285  37 13218  3
\end{verbatim}
\end{table}

\clearpage

\begin{deluxetable}{rrrrcl}
\tabletypesize{\normalsize}
\tablecaption{Description of BSCC data columns and range of data values.}
\tablewidth{0pt}
\tablehead{\colhead{BSCC  } & \colhead{min} & \colhead{max} & 
           \colhead{number} & \colhead{unit} & \colhead{description} \\
           \colhead{column} & \colhead{data} & \colhead{data} & 
           \colhead{zeros} & \colhead{} & \colhead{of data item} }
\startdata
 1 &   1491551& 1294725510 &         0& mas & RA (ICRS at mean  \\ 
 2 & 178689890&  319898157 &         0& mas & DEC  epoch of obs.) \\
 3 &         0&        999 &     54401& mas & scatter sigma RA*cosDec\\
 4 &         0&        999 &     89491& mas & scatter sigma Dec \\
 5 &         1&        400 &         0& mas & model sigma RA*cosDec\\
 6 &         2&        400 &         0& mas & model sigma Dec \\
 7 &         1&        288 &         0&     & total numb. images this star\\
 8 &         1&        283 &         0&     & numb. images used for mean pos\\
 9 &         0&         10 &  13711030&     & numb. images rejected outliers\\
10 &       132&       7321 &         0& 1/1000 yr & mean epoch [year-2000] \\
11 &         0&          4 &  12828714&     & largest double star flag \\ 
12 &         0&          1 &  13023146&     & bad pixel flag (1=bad, else 0)\\
13 &      7470&      20000 &         0& mmag& mean R model fit magnitude \\
14 &      $-$1&        900 &     22693& mmag& sigma R mag \\
15 &         0&        145 &    583043&     & numb. images used for R mag \\
16 &      6846&      20000 &         0& mmag& mean V model fit magnitude \\
17 &      $-$1&        900 &     17299& mmag& sigma V mag \\
18 &         0&        144 &   4311081&     & numb. images used for V mag \\
19 &         0& 1321535437 &    614483&     & 2MASS ID number \\
20 &      2069&      30000 &         0& mmag& 2MASS J magnitude \\
21 &      $-$1&        999 &         0& mmag& 2MASS sigma J magnitude \\
22 &       818&      30000 &         0& mmag& 2MASS H magnitude \\
23 &      $-$1&        999 &         0& mmag& 2MASS sigma H magnitude \\
24 &     $-$13&      30000 &         0& mmag& 2MASS K magnitude \\
25 &      $-$1&        999 &         0& mmag& 2MASS sigma K magnitude \\
\enddata
\tablecomments{Undefined photometric errors are set to $-1$,
               and a value of 30000 for a magnitude indicates
               ``no data".}
\end{deluxetable}

\end{document}